\documentstyle[11pt]{article}
\@addtoreset{equation}{section}
\def\theequation{\arabic{section}.\arabic{equation}}

\def\section{\@startsection{section}{1}{\z@}{3.5ex plus 1ex minus
   .2ex}{2.3ex plus .2ex}{\large\bf}}
   \def\thesection{\arabic{section}}

\def\appendix{\setcounter{section}{0}
        \def\thesection{Appendix\ \Alph{section}}
        \def\theequation{\Alph{section}.\arabic{equation}}}

\renewcommand{\@}[1]{\sqrt{#1}}

\def\be{\begin{eqnarray}}
\renewcommand{\le}[1]{\label{#1}\end{eqnarray}}
\def\ee{\end{eqnarray}}

\def\ffract#1#2{\raise .35 em\hbox{$\scriptstyle#1$}\kern-.25em/
\kern-.2em\lower .22 em \hbox{$\scriptstyle#2$}}

\begin{document}

\rightline{RI-06-04}
  \rm\large \vskip
0.9in
\begin{center}
{\LARGE On 't Hooft's S-matrix Ansatz for quantum black holes}
\end{center}
\vspace{1cm}

\begin{center}
{\large Giovanni Arcioni
\footnote{E-mail: {\tt
arcionig@phys.huji.ac.il}}
\\
\vskip 1truecm  {\it Racah Institute of Physics\\
The Hebrew University of Jerusalem\\
Jerusalem 91904, Israel.}}

\end{center}

\vspace{.5cm}

\begin{center}
\bf Abstract
\end{center}

\vspace{.5cm} The S-matrix Ansatz has been proposed by 't Hooft to
overcome difficulties and apparent contradictions of standard
quantum field theory close to the black hole horizon. In this
paper we revisit and explore some of its aspects. We start by
computing gravitational backreaction effects on the properties of
the Hawking radiation and explain why a more powerful formalism is
needed to encode them. We then use the map bulk-boundary fields to
investigate the nature of exchange algebras satisfied by operators
associated with ingoing and outgoing matter. We propose and
comment on some analogies between the non covariant form of the
S-matrix amplitude and liquid droplet physics to end up with
similarities with string theory amplitudes via an electrostatic
analogy. We finally recall the difficulties that one encounters
when trying to incorporate non linear gravity effects in 't
Hooft's  S-matrix and observe how the inclusion of higher order
derivatives might help in the black hole microstate counting.
\vspace{.5cm}

\newpage


\section{Introduction}


The standard derivation of the Hawking radiation of black holes
shows that they evaporate through (approximately) thermal
radiation \cite{Hawking}, \cite{predictability}. This implies the
well known problem of loss of information. In addition Hawking
radiation originates in very high frequencies vacuum modes at past
null infinity and gets to transplanckian energies near the
horizon.

An attempt to overcome all these problems has been proposed by 't
Hooft and goes under the name of S-matrix Ansatz \cite{gerard}. It
should also be a direct way to implement concretely the
holographic principle \cite{dimensional}, \cite{ologramma}. The
latter seems to provide the only remedy to deal with the formation
of bubbles hidden by the black hole horizon (which follows
unavoidably from General Relativity). This can be seen for
instance considering Schwarzschild black hole and working in the
volume gauge; one discovers that larger fractions of three-volume
will occupy the region beyond the horizon. It is therefore
reasonable to truncate the Hilbert space of any quantum theory one
has in mind to the horizon {\it surface} rather than to the volume
of the black hole.  This agrees with the fact that the black hole
entropy goes like an area, not like a volume. The latter
observation is normally assumed to be the starting point to
justify the holographic principle.

 In this paper we revisit the S-matrix Ansatz and explore some of
its implications. The paper is organized as follows: in Section
$2$ we review some aspects of the whole proposal and fix notations
and conventions. We comment in particular on the definition of the
horizon operators. In Section $3$ we examine and compute some
effects of the gravitational backreaction on the properties of the
Hawking radiation along the philosophy suggested by the S-matrix
Ansatz. In Section $4$ we specify the relation between bulk and
boundary fields and use it to study exchange algebras structures
which have appeared before in the literature. In Section $5$ we
propose and comment on some analogies between the non covariant
form of the S-matrix amplitude and liquid droplet physics. We also
point out similarities with string theory. In Section $6$ we
recall the difficulties encountered when attempting to incorporate
the non linearities of the gravitational force and observe how the
inclusion of higher order derivatives might help in the black hole
microstate counting. We finally summarize and give some conclusive
remarks in Section $6$.

\section{Review of the S-matrix Ansatz}

In this Section we revisit the S-matrix Ansatz proposal. The
interested reader can find a detailed discussion and related
references in the review of 't Hooft \cite{gerard}. Here we simply
recall and expand some points which are considered in the rest of
this paper and also fix notations and conventions.

The S-matrix Ansatz basically assumes that ``all physical
interaction processes that begin and end with free, stable
particles moving far apart in an asymptotically flat spacetime,
therefore {\it also} all those that involve the creation and
subsequent evaporation of a {\it black hole}, can be described by
one scattering matrix $S$ relating the asymptotic outgoing states
$\mid out >$ to the ingoing states $\mid in >$.

The starting point is to ${\it assume}$ a S-matrix amplitude $< in
\mid out>$ and compute then, including the effects of the
interactions $\delta_{in}$ and $\delta_{out}$, the neighboring
S-matrix elements $< out + \delta_{out} \mid in + \delta_{in}> $.

As a first step one considers gravitational interactions only. We
will see below the approximation in which the whole derivation is
carried out. For the moment just recall that the effects of these
interactions are basically described by gravitational shock waves:
incoming particles are strongly boosted as soon as they get close
to the black hole horizon and generate gravitational waves with
impulsive profile, typically a Dirac delta with support on a null
hypersurface. The net effect is a shift in the position of the
horizon and a shift of the geodesics of the outgoing particles.
Remarkably the shift can be computed explicitly \cite{dray}.

More properly one considers a factorization of the S-matrix
amplitude of the following form
 \be
 \label{splitting} S = S_{out} S_{hor} S_{in}
\ee
where $S_{in}$ is supposed to relate asymptotic in-states wave
packets to wave packets moving inwards very near the horizon. It
only describes what goes into the black hole. $S_{out}$ links wave
packets travelling outwards very near the horizon to asymptotic
out-states. It is supposed to describe all particles which come
out of the black hole and to be the time reversal of $S_{in}$.
$S_{hor}$ represents the non trivial part of the amplitude and
tells us how ingoing particles very near the horizon affect the
outgoing ones.

Let us observe that this splitting should in principle give no
complications \footnote{See however \cite{Itzhaki}. We would also
like to add that the location of the gluing of $S_{hor}$ with
$S_{out}$ is relevant to establish where the virtual fluctuations
due to gravitational polarization effects become Hawking
particles. } when considering the Rindler limit, i.e. a large mass
black hole. Actually particles spend on average a time $t_A \sim 4
M \ln M$ near the horizon before getting closer to it than the
Planck length and near horizon interactions are expected to take
place around this time scale. On the other hand, the average
evaporation time of the black hole is $t_B \sim {\cal O}( M^3)$.
It is reasonable to assume $t_B \gg t_A$ for large mass $M$ and
therefore one can approximate Kruskal coordinates with Rindler
coordinates near the horizon. In particular one can replace the
Kruskal angular coordinates $ \Omega = (\theta, \phi )$ with
Rindler coordinates $\tilde{x} =( x,y)$. We will use this
replacement in many points in the following.

If one considers the standard Penrose diagram of Schwarzschild
metric in Kruskal coordinates, 't Hooft's S-matrix puts into
communication, as a consequence of backreaction effects, region I,
III (i.e. the region out and in the black hole) with regions II
and IV (i.e. the in and out regions of the white hole). There
appears to be then a symmetry bewteen the black hole and the white
hole from the very beginning. Of course this is not standard,
since these two ``worlds" are normally disconnected.

We come now to another {\it not}-trivial feature of the S-matrix
Ansatz. One assumes that the ingoing and the outgoing states are
fully specified once the longitudinal momenta {\it only} $p_{in}
(\tilde{x})$ and $p_{out} (\tilde{x})$ are given as functions of
the transverse coordinates $\tilde{x}$. In other words, $\mid in
> = \mid p_{in} (\tilde{x})>$ and $\mid out > = \mid p_{out} (\tilde{x})
>$. This approximation corresponds to an eikonal
limit in which the transverse components $\Pi_{in} (\tilde{x})$
and $\Pi_{out} (\tilde{x})$ have been {\it neglected}. We discuss
the nature of this limit and its relation to the standard eikonal
resummation in Section $6$.

The relations of these momenta distributions with the components
of the stress energy tensors (in Rindler coordinates) are
 \be \label{momenti}
p_{in}(\tilde{x}) & = & \int T_{++} (x^+, x^-, \tilde{x}) dx^+
\mid_{x^- = 0}
\\
p_{out} (\tilde{x}) & = & -\int T_{--} (x^+, x^-, \tilde{x}) dx^-
\mid_{x^+ = 0} \\
\Pi_{in}^a  (\tilde{x})& = & \int T_{a +} (x^+, 0, \tilde{x} ) dx^+ \\
\Pi_{out}^a (\tilde{x}) & = & - \int T_{a -} (x^-, 0, \tilde{x} )
dx^-
 \ee
with $a=1,2$ running over the transverse coordinates.

One can go then to the dual position operators via (functional)
Fourier transform. Consider for instance the in-state $\mid p_{in}
(\tilde{x}) >$ (with eigenvalue $p_{in} (\tilde{x})$). Its
conjugate will be given by \be \mid x_{in} (\tilde{x}) > =  \int
[Dp_{in}] e^{-i \int d \tilde{x} p_{in} (\tilde{x})
x_{in}(\tilde{x})} \mid p_{in} (\tilde{x}) >
 \ee
with
 \be < x_{in} (\tilde{x}) \mid p_{in} (\tilde{x}) > = e^{i
\int d \tilde{x} p_{in} (\tilde{x}) x_{in} (\tilde{x}) }
 \ee

One assumes then standard canonical commutation relation for the
in and out bases respectively, i.e. \be [ p_{in} (\tilde{x}),
x_{in} ( \tilde{x}')]= -i \delta^2 (\tilde{x} - \tilde{x}')
 \ee
\be
 [ p_{in} (\tilde{x}), p_{in} ( \tilde{x}')] = 0 , [ x_{in}
(\tilde{x}), x_{in} ( \tilde{x}')] = 0
 \ee
and analogue relations
for the out sector.

Non trivial relations between in and out bases follow from taking
into account gravitational backreaction effects. Explicitly one
gets \cite{gerard}, \cite{dray}
\be \label{relazioni} x_{in} (
\tilde{x}) & = & - \int d \tilde{x}' f(\tilde{x} -
\tilde{x}') p_{out} (\tilde{x}')\\
x_{out} (\tilde{x}) & = & \int d \tilde{x}' f( \tilde{x} -
\tilde{x}') p_{in} (\tilde{x}')
\ee
 where $f( \tilde{x} -
\tilde{x}') $ is the shift due to gravitational shock waves which
can be exactly computed. This shift goes like $\ln (\tilde{x} -
\tilde{x}')$ and therefore diverges for small angular separations
(where by small we mean of Planck length order). We have assumed,
however, as a first approximation, to neglect the transverse
components of the momentum and this means that the resolution will
be always bigger than the Planck length.

What is {\it not} trivial is the algebra one gets once
(\ref{relazioni}) are taken into account. Indeed one easily gets
\be
\label{xinxout}
 [ x_{out} (\tilde{x}), x_{in} (\tilde{x}')]  =
- i f(\tilde{x} - \tilde{x}'). \ee
 \be
 [ p_{in} (\tilde{x}) , p_{out} (\tilde{x}')]   =  - i \tilde{
\partial }^2 \delta^2 ( \tilde{x} -\tilde{x}')
 \ee

These relations tell us that there are non trivial correlations
between ingoing and outgoing matter.

Note however the following important point: the operators
$x_{in(out)} (\tilde{x})$ refer to the horizon shape {\it not} to
single particles. Thinking therefore \be [ x_{in,i} (\tilde{x}) ,
x_{out,j} (\tilde{x}')] = - i f (\tilde{x} - \tilde{x}')
\delta_{ij} \ee
 i.e. to single
particles labelled by indices $i,j$ would be wrong. Actually all
ingoing particles interact (gravitationally) with all outgoing
particles. For instance, if one has two particles $1,2$ then
$x_{in,1} (\tilde{x}) - x_{out,2} (\tilde{x})$ would commute with
everything and this would imply loss of information, which we want
to avoid from the beginning. Several recent papers on non
commutative quantum field theory pointed out a link between Moyal
deformed algebras and 't Hooft's algebras. However in the former
case one is referring to particle coordinates.

Therefore one has to interpret $x_{in(out)} (\tilde{x})$ as
 \be
\label{somma} x_{in(out)} (\tilde{x}) = \sum_i x_{in(out),i}
(\tilde{x})
 \ee
where the sum is over all ingoing (outgoing) particles at the {\it
same} transverse position $\tilde{x}$.

As a consequence a single real number encodes the locations of
{\it all} particles at a given $\tilde{x}$. All the coordinates
are in this way at the right side of the horizon regardless how we
shift the total $x_{in (out)} (\tilde{x})$ and given (\ref{somma})
one can reobtain all the entries in the sum in an unambiguous way
\cite{gerard}. From a dual perspective, the momenta distributions
$p_{in (out)} (\tilde{x})$ correspond to the momenta of all
particles. This is similar, as pointed out by 't Hooft and
Susskind to what happens in QCD parton models. This non
conventional distribution of particles and their not defined
locations suggest non trivial physics, which should emerge when
one zooms to planckian resolution in the transverse coordinates.

Using then the previous relations one can show in a
straightforward way that the S-matrix amplitude is given by
\be
\label{pinout} < p_{out} (\tilde{x}) \mid p_{in} (\tilde{x}) > =
{\cal N} \exp [ -i \int d^2 \tilde{x} d^2 \tilde{x}' p_{in}
(\tilde{x}') f(\tilde{x} - \tilde{x}') p_{out} (\tilde{x}) ]
 \ee
where ${\cal N}$ is a normalization factor which is supposed to be
fixed by unitarity. Manipulating further expression (\ref{pinout})
one gets
\be
\label{localenoncovariante}< p_{out} (\tilde{x}) \mid
p_{in} ( \tilde{x} ) >
 = {\cal N} \int
[D x_{in}(\tilde{x})] \int [ Dx_{out} (\tilde{x})]  \times \\
\times e^{i \int d^2 \tilde{x} ( - \tilde{\partial} x_{out}
(\tilde{x}) \tilde{\partial} x_{in} (\tilde{x}) + p_{in}
(\tilde{x}) x_{in} (\tilde{x}) - p_{out} (\tilde{x}) x_{out}
(\tilde{x})
      ) }
 \ee
Defining then $X^\mu =(x_{in},x_{out} )$ and  $p^\mu  = ( p_{in},
- p_{out})$ one can also rewrite the amplitude in a covariant way
 \be  \label{formalocale}
 < in \mid out> \sim \int d \tilde{x} ( \frac{1}{2} (
\tilde{\partial} X^\mu )^2 + p^\mu X_{\mu})
   \ee
Interestingly, if one imposes a static gauge on the {\it
transverse} coordinates one gets the string theory action of
Veneziano amplitudes (though with an imaginary tension despite the
euclidean worldsheet). This is in turn suspected to give
Nambu-Goto action once transverse components are taken into
account. Even so, however, one does {\it not} recover a finite
number of states.

The final target of the whole S-matrix Ansatz program would be
thus to derive the dynamics of a {\it finite} number of degrees of
freedom living on the horizon. The inclusion of the transverse
components of the momentum is a first step in this direction.
Further degrees of freedom should emerge from the inclusion of the
remaining interactions, once again when one probes Planck distance
scales in the transverse plane.

Thinking along similar lines 't Hooft has derived a {\it
covariant} algebra for the horizon degrees of freedom. The
building blocks are chosen to be the orientation tensors of the
horizon \be \label{orientation}
 W_{\mu \nu} =
\epsilon^{ij} \partial_i X_\mu
\partial_j X_\nu
  \ee
which have been shown to satisfy the covariant (local) algebra
 \be
\label{orienalgebra} [ W^{ \mu \alpha} ( \tilde{x}) , W^{\mu
\beta} (\tilde{x} ') ] = \frac{1}{2} \delta^2 (\tilde{x} -
\tilde{x}') \epsilon^{\alpha \beta \mu \nu} W^{\mu \nu}
(\tilde{x})
 \ee
  This algebra has been derived {\it neglecting}
second order derivatives in the embedding coordinates $X^{\mu}$
describing the horizon fluctuations. Despite covariant, some of
its generators are not hermitian. This means at the end of the day
that that one does not end up with a finite number of states. We
return on these aspects in Section $6$.


\section{Gravitational backreaction and Hawking radiation}


In this Section we start to examine the effects of the
gravitational backreaction on the Hawking radiation. As we will
see both the spectrum and the correlators display slight deviation
from pure thermal radiation but these effects are transient and
one needs a more powerful formalism to encode them properly. This
is what the S-matrix Ansatz is indeed supposed to provide as we
will discuss at the end of this Section.

We consider then for simplicity a null spherically symmetric shell
of matter with energy $\delta M$ which falls into a Schwarzschild
black hole of mass $M$ (with $\delta M \ll M $) at some late
advanced time\footnote{As usual, $u= t-r^*$, $v=t + r^*$  and
$r^*$ is the tortoise coordinate $r^* = r + 2M \ln (r/(2M)-1)$.}
$v_1$. The formation time of the horizon $v_0$ is thus shifted to
a slightly earlier time $v_0 \rightarrow v_0 + \delta v_0$ and
\cite{vvexchange}
  \be \label{shiftv}
   \delta v_0= -4
\delta M \exp \left( \frac{v_0 - v_1}{4M} \right)
\ee
 A light ray that
originally would have reached the outside observer at some
retarded time $u$ will arrive with a time delay $\delta u (u)$
 \be
\label{shift2} \delta u (u) = -4M \ln \left( 1+ \frac{\delta
v_0}{4M} \exp \left( \frac{u-v_0}{4M} \right) \right)
 \ee
  Notice that although the shift (\ref{shiftv})
 is small, it can have relevant physical effects on the outgoing modes:
 indeed, one sees from (\ref{shift2}) that the time delay $\delta u
 (u)$ diverges after a {\it finite} time $\bar{u}$
\be \label{divergenzau} \bar{u} - v_0 \sim -4M \ln \left(
\frac{\mid \delta v_0 \mid}{4M} \right)
 \ee

In what follows we examine the effects of the shift
(\ref{shiftv}), i.e. a classical backreaction effect,  on the
properties of the outgoing Hawking radiation. Since one is not
able to perform a full dynamical calculation, one repeats the same
steps of Hawking derivation, this time however taking into account
the shift (\ref{shiftv}) in the outgoing modes. As a consequence,
one gets corrected Bogoliubov coefficients
\be
\label{corrected}
\left\{ \begin{array}{ll}
   \alpha_{\omega \omega'}^{corr} \\
   \beta_{\omega \omega'}^{corr}
 \end{array} \right\}
 =\frac{1}{2\pi} \sqrt{\frac{\omega '}{\omega}}
\int_{- \infty}^{v_0} dv \exp (\mp i \omega ' v ) \exp (i \omega
(u + \delta u))
\ee

Here and in the following we use the abbreviation $corr$ to
remember that the coefficients which are calculated are corrected
due to the presence of the shift $\delta u (u)$. One is thus
computing the effect of the incoming shell of matter on the
outgoing Hawking radiation.

Using the well known diffeomorphism found by Hawking
\cite{Hawking} relating advanced and retarded coordinates

\be \label{uvdiff} u(v)= v_0 -4M \ln \left( \frac{v_0 -v}{4M}
\right)
\ee
 together with (\ref{shift2}) one gets
 \be
\label{corretti} \left\{ \begin{array}{ll}
   \alpha_{\omega \omega'}^{corr} \\
   \beta_{\omega \omega'}^{corr}
 \end{array} \right\}
= \frac{1}{ 2\pi } \sqrt{\frac{\omega '}{\omega}} e^{i \omega v_0
+ 4 M i \omega \ln 4M} \int_{- \infty}^{v_0} dv e^{\mp i \omega '
v } (v_0 - v -D)^{-4 M i \omega}
 \ee
 where $D$ is given by
 \be
  D =4 \delta M \exp
\left( \frac{v_0 -v_1 }{4M} \right)
\ee
 Putting now $(v_0 -v -D) =
x$ one has \be \left\{ \begin{array}{ll}
   \alpha_{\omega \omega'}^{corr} \\
   \beta_{\omega \omega'}^{corr}
 \end{array} \right\}
 =  \frac{1}{ 2\pi } \sqrt{\frac{\omega '}{\omega}}
e^{i \omega v_0 + 4 M i \omega \ln 4M \mp i \omega ' v_0 \pm i
\omega ' D} \int_{- D}^{+ \infty} dx e^{\pm i \omega ' x} x^{- 4 M
i \omega}
 \ee
 Recalling the definition of incomplete Gamma function
 \footnote{
 The incomplete Gamma function is defined as \cite{tavole}
\be \label{defgammaincompleta} \Gamma (\alpha , z) =
\int_z^{\infty} dt e^{-t} t^{\alpha -1} \ee}
 one finally obtains
\be
\label{finalecorretti}
 \left\{ \begin{array}{ll}
   \alpha_{\omega \omega'}^{corr} \\
   \beta_{\omega \omega'}^{corr}
 \end{array} \right\}
 = \pm \frac{i}{2 \pi \sqrt{ \omega \omega '}} e^{i (\omega \mp \omega ') v_0
  \pm i \omega ' D + 4M i \omega \ln 4 M \omega ' \pm 2 \pi M
  \omega} \Gamma ( 1 - 4 M i \omega , \pm i \omega ' D)
 \ee
  It can be easily checked
that in the limit $D \rightarrow 0$ (i.e. no back-reaction) one
recovers the same of \cite{Hawking}.

In the present case, however, the Bogoliubov coefficients depend
not only on the usual parameters like the black hole mass $M$ and
the time formation of the horizon $v_0$ but {\it also} on the
parameters of the infalling shell, namely the energy $\delta M$
and the time $v_1$. This suggests that the evaporation process
becomes highly {\it dynamical} if one takes into account the
backreaction, in accordance with 't Hooft's scenario.

Consider now the effects of this shift on the thermal properties
of the Hawking radiation starting from the the spectrum. The
Bogoliubov coefficients which give the main contribution are those
with very large values of the frequency $\omega '$. Actually if an
outgoing wave reaches infinity at late times with a finite
frequency $\omega$ then the incoming waves which contribute to it
at early times must have a very large frequency $\omega '$ due to
the large red-shift induced by the geometry of the black hole.

One is thus interested in the large $\omega ' $ limit. Using the
asymptotic expansion of the incomplete Gamma function
\footnote{One has for $\mid z \mid \rightarrow \infty$ \be
\label{asymptgammalarge} \Gamma (\alpha ,z) \sim z^{\alpha -1}
e^{-z} \left( 1+ \frac{\alpha -1}{z} + O(z^{-2}) \right)
 \ee
  This
holds properly for $\mid Arg (z) \mid < \pi/2$. In our case $z=
\pm i \omega ' D $  so a small real part $\epsilon$ has been added
as usual to regularize producing  a subleading $\epsilon$
dependent factor which can then be removed.} one gets at leading
order \be \label{altefrequenze} \mid \frac{\alpha_{\omega \omega '
}^{corr}}{\beta_{\omega \omega'}^{corr}} \mid^2 \sim \exp (8 \pi M
\omega ) \ee
 times corrections organized in power of $1 / \omega '$.

One therefore gets small deviations from the Planckian spectrum,
which {\it also} contain explicitly the frequency $\omega'$. These
corrections, however, modify the Planckian nature of the spectrum
only for a very short time.

An analogy can be made along these lines: suppose we consider some
water at equilibrium in a bowl and we send in a microscopic body
with high energy. The water will slightly change its temperature
and thermodynamic properties and will return to equilibrium almost
immediately. The quantum state of the system has changed however
and one should have a powerful formalism to detect these changes.

This precisely led 't Hooft to propose the S-matrix Ansatz as a
possible mechanism to encode all the changes of the black hole
quantum state.

It is interesting to consider the stochastic properties of the
outgoing radiation as well, i.e. the structure of the n-th order
correlation functions. Indeed the radiation could have gaussian
correlators without having a planckian spectrum or vice versa
\cite{sciama}, something which happens for instance in quantum
optics.

In the present case this means that one has to calculate not only
the expectation number of the emitted particles $<N_\omega>$ (i.e.
the spectrum)  but also $<N_\omega^2>$, $<N_\omega^3>$ and so on
\cite{predictability}.

Consider for instance $<N_\omega^2>$. This is given by
 \be \label{correlatore}
 <N_\omega^2>= < in \mid
(b_\omega^\dagger b_\omega )(b_\omega^\dagger b_\omega) \mid in
>
\ee
 With the aid of the Bogoliubov transformations one gets
 \be \label{2punti}
< N_\omega^2 > = < N_\omega > + 2 (< N_\omega >)^2 + \sum_i
\alpha_{ji} \beta_{ji} \sum_k \alpha_{jk}^* \beta_{jk}^*
\ee

Following \cite{predictability} one has therefore to evaluate
expressions of this form \be \label{nonvanishing} \int
\alpha_{\omega_1 \omega'} \beta_{\omega_2 \omega'} , \int
\alpha_{\omega_1 \omega'}^* \beta_{\omega_2 \omega'}^*
 \ee
  Using the standard Bogoliubov
coefficients these integrals vanish (since proportional to $\delta
(\omega_1 + \omega_2)$ and $\omega$ is positive). One has then a
gaussian correlator, as expected in the case of thermal radiation.
On the contrary, using the corrected Bogoliubov coefficients
(\ref{finalecorretti}) these expressions do not vanish;
interestingly they have the {\it same} order of divergence
(logarithmic) as the two previous terms on the r.h.s of
(\ref{2punti}).

This is again in agreement with 't Hooft's scenario, implying non
trivial correlations between ingoing and outgoing matter. Still,
however, these correlations disappear after a very short time as
just pointed out. All these transient deviations, however, suggest
that the resulting Hilbert space of the system is not any more the
tensor product of in and out (w.r.t the black hole) Hilbert
spaces.

\section{Holography at work: bulk-boundary fields and exchange algebras}
The computations of the previous Section suggests that the
outgoing matter is thus related to the ingoing one. This fact has
been used by 't Hooft to derive a non trivial algebra
(\ref{xinxout}) satisfied by ``horizon operators", whose (quite
not conventional) definition has been recalled in Section $2$.
Following the most recent developments of the holographic
principle, these are {\it boundary} fields, namely fields which
live on the boundary ``screen"-the horizon-where the holographic
theory is supposed to live (See for instance the recent review
\cite{janliatasad}).

In \cite{vvexchange}, however, an interesting exchange algebra
satisfied by ingoing and outgoing fields was derived too. The
fields which enter in this algebra, however, are now {\it bulk}
fields and describe the higher dimensional physical world which is
supposed to be encoded in the boundary description.

Consider indeed a scalar field $\phi$ propagating on a
Schwarzschild background. One can easily show that close to the
horizon the solutions of the Klein-Gordon equation split in an
ingoing $\phi_{in}$ and an outgoing component $\phi_{out}$ and it
was proposed in \cite{vvexchange} that once {\it quantum}
backreaction effects are taken into account one gets to a non
trivial exchange algebra of the form \be
 \label{algebradiscambio}
\phi_{out}(u , \Omega) \phi_{in} (v, \Omega ') = \exp \left( - 16
\pi i f(\Omega, \Omega ' ) e^{\frac{u-v}{4M}} \partial_u
\partial_v \right) \phi_{in} (v,\Omega ') \phi_{out} (u, \Omega)
\ee
 where $f(\Omega, \Omega ')$ is the shift due to the
shock wave discussed before.

The derivation of this algebra, however, is quite not trivial and
one has to make {\it several} assumptions in order to get to this
result. One of the curious things already noticed in
\cite{vvexchange} is that the final form is symmetric in the in
and out fields, while the derivation does not treat them in this
way. In addition, one has to promote to operator the formation
time of the horizon $v_0$. One assumes then a resolution in the
transverse angular coordinates $\Omega , \Omega '$ bigger than the
Planck length (i.e. small transverse momenta); this last
assumption, a sort of eikonal limit, is the same made by t'Hooft
but in addition one has to take $v$ sufficiently bigger than
$v_0$.

In this Section we would like to derive a similar result in a more
direct way, using the prescription given by 't Hooft to map bulk
fields into boundary ones.

The ``holographic map" bulk-boundary fields proposed by 't Hooft
acts in this way: take a field at some point P in the bulk and a
point Q on the horizon. If the proper distance PQ is finite, then
the bulk field will carry the same representations of the Hilbert
space defined on the horizon. 't Hooft derives his S-matrix in the
Rindler limit (which is assumed in the next considerations), so in
that case all these proper distances are finite. The whole region
is thus described holographically by the horizon fields. In other
words, the Hilbert space associated with the degrees of freedom
living on the horizon is supposed to encode the whole universe
outside the black hole.

To fix ideas, consider for instance a bulk scalar field satisfying
the Klein Gordon equation. In the limit of zero mass and zero
transverse momenta, the bulk field $\phi(x^+ , x^- , \tilde{x})$
splits into an ingoing component $\phi_{in} (x^+,\tilde{x} )$ and
an outgoing component $\phi_{out } ( x^- ,\tilde{x})$, where $x^+
, x^-$ are the usual light cone coordinates and $\tilde{x}$
represents the transverse coordinates \footnote{Recall that $x^+ =
x+t$ and $x^- = x-t$ and $k^0 = \mid k_x  \mid $ in the present
situation.}.

Using mixed Fourier transform 't Hooft has shown that up to a good
approximation
\be
 \label{bulkin} \phi_{in} ( x^+ , \tilde{x}) \sim
\int d k_x \exp ( i k_x x^+ - i k_x x^+ (\tilde{x}) )
\ee
  On the r.h.s one has $x^+ (\tilde{x})$,
the {\it horizon} operator (which we denoted as
$x_{in}(\tilde{x})$ in Section $2$). A similar, completely {\it
symmetric} expression holds naturally for the outgoing field $x^-
(\tilde{x})$ (denoted by $x_{out} (\tilde{x})$ again in Section
$2$).

Note the non trivial relation between the bulk field $\phi_{in}$
 and the horizon operator $x^+ (\tilde{x})$. Similar complicated
mappings are to be expected from the holographic principle, where
a lower dimensional theory encodes higher dimensional bulk data.
See for instance \cite{tom} in AdS/CFT correspondence set up. The
't Hooft eikonal limit can be rephrased also in terms of a $2+2$
splitting of Einstein gravity with two coupling constants
\cite{scatteringverlinde}. In the case of non vanishing
cosmological constant the corresponding algebras become in
principle much more complicated \cite{our}.

We have used these expressions to obtain the commutator between
ingoing and outgoing bulk fields. The computation is
straightforward and one gets (using (\ref{splitting}))
 \be \label{veraalgebra} [
\phi_{in} (x^+ ,\tilde{x}) , \phi_{out} ( x^- , \tilde{x}') ] = i
f (\tilde{x}- \tilde{x}')
\partial_+ \phi_{in} ( x^+ ,\tilde{x})
\partial_- \phi_{out }(x^- , \tilde{x}')
\ee This non local algebra shows explicitly the complicated
relation between in and out bulk fields, which would clearly
commute in the case of no backreaction.

 From this one has in an obvious way
  \be
\label{similarvv} \phi_{out} (x^- , \tilde{x}) \phi_{in} (x^+ ,
\tilde{x}') = (1- i f(\tilde{x}-\tilde{x}') \partial_{x^{+}}
\partial_{x^{-}}) \phi_{in} (x^+ ,\tilde{x}') \phi_{out} (x^-, \tilde{x})
\ee

Provided that, as said, the angular separation in the transverse
coordinates is bigger than the Planck scale one is thus invited to
exponentiate the r.h.s. obtaining thus precisely(!) the Verlinde
exchange algebra (\ref{algebradiscambio}) though in the Rindler
limit.

The relation (\ref{veraalgebra}) seems however to us most
fundamental. The exponentiation is indeed only formally justified
in \cite{vvexchange} along the lines of \cite{abarbanel}, where it
was shown how to re-sum explicitly ladder diagrams in the eikonal
approximation. In this situation, however, it is not completely
clear to which Feynman diagrams-rules one is referring.

In addition the in-out symmetry of 't Hooft's S-matrix Ansatz is
present by definition from the very beginning. As we saw in
Section $1$ the splitting (\ref{splitting}) considers the in and
out sectors on equal footing. On the other hand, the symmetry in
the algebra (\ref{algebradiscambio}) is only recovered a {\it
posteriori}. Actually $\phi_{in}$ is supposed to evolve via a sort
of Moller operator to $\phi_{out}$ or $\phi_{hor}$ if $v < v_0$ or
$v > v_0$ respectively. Notice also that in \cite{svv} it is
postulated that the energy $E$ satisfies
 \be
 [E,v_0]=i
 \ee
It  immediately follows (using  standard Bogoliubov coefficients)
 \be [E,\alpha_{\omega \omega'}]=
 (\omega' - \omega)
\alpha_{\omega \omega'} ,[E,\beta_{\omega \omega'}] = (\omega' +
\omega) \beta_{\omega \omega'}
 \ee
and one can give a physical interpretation to the Bogoliubov
coefficients since they carry energy $E$ equal to $\omega'
-\omega$ and $\omega' + \omega$ respectively. In the presence of
backreaction, however, if one uses the corrected Bogoliubov
coefficients we computed before this relation is not true any
more.

\section{The horizon as a membrane}

As shown in Section $2$, by mean of Fourier transformation one can
recast (\ref{pinout}) in the local form (\ref{formalocale}). In
this Section, however, we are going to consider the (not trivial)
expression (\ref{pinout}) itself describing small fluctuations of
the horizon around spherical symmetry.

We first use an analogy with liquid droplets, assimilating the
black hole horizon to a droplet interface, to show how
(\ref{pinout}) comes out in a natural way by considering small
deformations of the droplet itself. We point out then where we
believe the analogy fails and list arguments which suggest to
interpret the horizon as a fluctuating membrane, in agreement with
standard proposals \cite{uglum}, \cite{Price}. We finally use an
analogy with two dimensional electrostatic to interpret again
(\ref{pinout}) this time from a stringy perspective.

The analogy with liquid droplets as been pointed out in a series
of papers by Kastrup \cite{kastrup}, \cite{z2}. Indeed, if one
assumes- following Bekenstein old suggestion \cite{jacob1}-a
discrete spectrum for the black hole with energy levels $E_n$
quantized as $E_n = \sigma \sqrt{n} E_{pl}$, $(\sigma = O(1))$,
and a degeneracy $g^n$ (with $g$ positive), the canonical
partition function is clearly given by \be
\label{partitiondroplet} Z= \sum_{n=0}^\infty e^{nt} e^{- \sqrt{n}
x } \ee
 where $t= \ln g$
and $x= \beta \sigma
 E_{Pl}$. Kastrup has noticed that this partition function is of
 the same form of the one describing liquid droplets: namely the
term in the exponent going like $n$ is a bulk energy while the one
going like $\sqrt{n}$ is a surface energy contribution. Using
analytic continuation arguments along the lines of the elegant
work of Langer \cite{langer}, he also explicitly computed this
partition function and found, interestingly, that its imaginary
part for $g>1$ reproduces the standard thermodynamics properties
of the Schwarzschild black hole (the result holds in any
dimension).

Let us then temporary use this analogy to examine the small
deformations of a spherical droplet. Recall that one of the
features of the droplets physics is that once the surface tension
$\tau$ is given, all the properties depend on the geometry of the
interface which separate the droplet from the bulk.

Suppose then that we want to consider small fluctuations of the
droplet around spherical symmetry. An internal pressure $p_{in}$
and an external pressure $p_{out}$ act on the droplet. $p_{in}$
will be given by $p_{in}^{eq} + \delta p_{in} (\Omega)$ where the
first contribution stays for the internal pressure at equilibrium
while the second contribution describes the small change in the
internal pressure due to the perturbation (we allow an {\it
angular} dependence precisely as in 't Hooft). Analogue splitting
holds for the external pressure $p_{out}$.

At equilibrium Laplace formula must hold \cite{LL}, namely
\be
\label{laplaceformula} \tau \left( \frac{1}{R_1}+ \frac{1}{R_2}
\right) = p_{in} - p_{out}
\ee

where $R_1$ and $R_2$ are the principal radii of curvature of the
droplet. In the absence of perturbations, one obtains $p_{in}^{eq}
- p_{out}^{eq} = (2 \tau)/r$, where r is the radius of the droplet
at equilibrium.

For small fluctuations $r \rightarrow r + \delta r (\Omega)$, at
first order, one gets easily
\be
\label{dropletdeformed} (\nabla
+2 ) \delta r (\Omega) = \delta p_{out}(\Omega) - \delta p_{in}
(\Omega)
\ee
 where $\nabla$ is the laplacian on the sphere. At
this point it is convenient due to the symmetry of the problem to
expand in spherical harmonics $\delta r (\Omega)$, $ \delta
p_{out}(\Omega)$ and $ \delta p_{in} (\Omega)$ to get
\be
 \label{canali} \delta R^{lm} (\Omega) = \frac{\delta
p_{in}^{lm}(\Omega) - \delta p_{out}^{lm} (\Omega) }{l(l+1)-2}
\ee

We consider now-in analogy with the 't Hooft model of the horizon-
the deformation due to a force localized in a point, which is then
described by a Green function $f(\Omega,\Omega')$. The latter will
satisfy \be \label{dropletshift}
 (\nabla +2) f( \Omega, \Omega') = - \delta^{(2)}
(\Omega, \Omega') = - \sum_{l,m} Y_{lm}^* (\Omega) Y_{lm} (\Omega
')\ee
 The sum
however has to reflect in this case the physics of the droplet:
the $l=1$ component is source of divergence as can been seen from
(\ref{canali}) and has to be removed. It corresponds to an
infinite translation. If one assumes incompressibility too, then
the $l=0$ component has to be removed too to keep the internal
volume of the droplet unchanged (we have checked however that the
small angle behavior of the Green function does not change even
when incompressibility is not imposed).

Assuming thus $l \geq 2$ in (\ref{dropletshift}) one gets
\footnote{We used the series in 8.92 of \cite{tavole}.}
 \be
 \label{greenfunction} f(\theta)
\sim \frac{1}{2} + \frac{2}{3} \cos ( \theta ) + \cos (\theta) \ln
(\sin^2 (\frac{\theta}{2}))
 \ee

Therefore the energy $U$ (i.e. work) required to deform the
droplet is given by
\be \label{energydeformation} U \sim \int d
\Omega d \Omega ' \delta p_{in} (\Omega) f(\Omega, \Omega' )
\delta p_{out} (\Omega ')
\ee

We notice that assimilating the black hole horizon to a droplet
interface one obtains basically the same expression of 't Hooft
and even if the form of the shift (\ref{greenfunction}) is
different, the small angle behavior is {\it again} logarithmical
as in 't Hooft model.

Black hole physics is however another story. We can point out the
differences starting by examining 't Hooft derivation of
(\ref{pinout}). Before doing this, however, recall that the
expression appearing in the exponent of the partition function
(\ref{partitiondroplet}) holds in general for large radius
droplets, corresponding to compact clusters associated with low
temperature physics. For smaller radius droplets, on the other
hand, one gets ramified droplets and hybrid structures, which
dominate the high temperature phase and {\it have} to be summed in
the partition function \cite{Domb}, \cite{Binder}.

Despite the analogy, however, expression (\ref{pinout}) derived by
't Hooft contains inputs peculiar to the black hole physics under
consideration. From the shock waves analysis, indeed, one has not
trivial relations among the longitudinal components of ingoing and
outgoing position-momentum operators \be
\partial_{\tilde{x}}^2 x_{in(out)} (\tilde{x}) = \mp p_{out(in)} (\tilde{x})
\ee Therefore our amplitude (\ref{pinout}) can be also rewritten
(formally) as (all fields depend on transverse coordinates
$\tilde{x}$
 \be \label{intermediate}
< p_{in} ( \tilde{x}) \mid p_{out} ( \tilde{x} ') > =  \int
[Dx_{in} ]
 \delta (\partial^2 x_{in} + p_{in}  )
e^{-i \int p_{out} x_{in}}
 \ee
and an analogue integration can be done for the other longitudinal
position operator giving
\be
< p_{in} ( \tilde{x}) \mid p_{out} (
\tilde{x} ') > = \int [Dx_{in}] [Dx_{out}] e^{-i \int \partial
x_{in}
\partial x_{out} + p_{in} x_{out} - p_{out} x_{in} }
 \ee
This is clearly a membrane like expression as recalled before. One
starts from (\ref{pinout}) and then Fourier transform to get
(\ref{localenoncovariante}). Here we just showed a different path
to get to the same result. The intermediate step
(\ref{intermediate}) shows the typical form of a path integral
over a membrane where the delta takes into account all possible
(in general not trivial) constraints imposed on the membrane
dynamics.

On general grounds, indeed, the membrane properties are different
from the ones of the droplet interface. Actually an interface
normally means a boundary between two phases whose fluctuations
can be studied by method adapted from equilibrium critical
phenomena. The statistical mechanics is normally controlled by the
surface tension. Membranes, on the other hand, are composed of
molecules {\it different} from the medium in which they are
embedded, and they need not separate two distinct phases. They
have in general a richer internal structure.

All of this is in agreement with 't Hooft picture, where once the
transverse momenta {\it and} the non gravitational interactions
are taken into account they are supposed to generate additional
degrees of freedom living on the horizon. Despite the similarity
with the energy for small deformations around spherical symmetry,
it is then clear that the black hole horizon cannot be simply an
interface.

Similar considerations hold in the case of the stretched horizon
picture \cite{uglum}, though one has in this case a different set
up. One is indeed interpreting the horizon as a $2+1$ dimensional
membrane embedded in spacetime (more precisely a {\it timelike}
hyper-surface). Therefore one looses the symmetry between in and
out states. As a matter of fact, if one adopt the complementarity
principle, from the point of view of the {\it static} external
observer the horizon is a quantized membrane with its own degrees
of freedom. One is therefore forced to use static Schwarzschild
coordinates in this case. The fact that the stretched horizon has
codimension one while the spacelike horizon two may have other
important consequences for the counting of states, as we will see
in Section $6$.

The expressions (\ref{pinout}) and (\ref{dropletshift}) have also
another interesting analogy. They resemble the two dimensional
electrostatic situation in which (\ref{pinout}) can be interpreted
as a scalar potential round a point charge. We want to use this
analogy now to make a link with string theory amplitudes.

Suppose we start from a four point function, (below we comment on
the n-point case). Imagine two ingoing and two outgoing massless
particles as external lines entering an electric circular circuit.
Let the coordinate along the circle be the Feynman parameter $x$;
the typical result of a one loop integration produces \be \int_0^1
\frac{dx}{x(1-x)} \times ext.insertions.
 \ee
The contribution of the external lines is again of the form
(\ref{pinout}). The leading contribution from the shift is as we
said logarithmic and now the "transverse coordinates" are replaced
by $x$. Suppose we insert the four massless particles at points
$0, x, 1, \infty$ (this choice will become clear in a moment). The
final result is then \be
 \int_0^1 \frac{dx}{x(1-x)}
 \exp ( -2 p^1  p^2 \ln (x) - 2 p^1 p^3 \ln (1-x))
    \ee
Use now the Mandelstam variables $s,t$. One gets
\be
\label{venezianodue} \int_0^1 \frac{dx}{x(1-x)}
 \exp (-s \ln (x) - t \ln (1-x))     \ee
which in turns is nothing than
 \be \label{venezianoformula}
\int_0^1 dx x^{-s-1} (1-x)^{-t-1} \ee and this is the celebrated
Veneziano formula!

Recall that if one considers a n-point function for massless
particles scattering in the eikonal limit (i.e. once again large s
and small t Mandelstam parameters) we have been discussing, one
can show \cite{poles} that the total transition amplitude nicely
factorizes into the product of two particle amplitudes of the form
(\ref{pinout}). To fix ideas consider for instance a 6-point
function. One has
\be
< k^1, k^2, k^3 \mid p^1, p^2, p^3 > & = &
\int dx^1 dx^2 dx^3
 <k^{(1)},k^{(2)} \mid x^{(1)}, x^{(2)} > \times \\
 & \times & <x^{(2)},k^{(3)} \mid
p^{(2)}, x^{(3)} > <x^{(1)},x^{(3)} \mid p^{(1)}, p^{(3)} > \ee
where $k,p$ stay for incoming and outgoing momenta respectively of
the particles $1,2,3$. We are currently trying to understand what
these generalizations correspond from stringy perspective to see
if more information can be extracted \cite{mio}.

\section{On the inclusion of transverse momenta}

As recalled in Section 1, the horizon operator algebra once
written in a covariant way should give the correct counting of the
black hole microstates. Promoting the orientation tensors
(\ref{orientation}) of the horizon (see additional comments below
in this Section) to operators, 't Hooft has obtained a covariant
algebra (i.e. Lorentz invariant) and the generators of these
algebra are supposed to correspond to different cell-domains
defined on the horizon. Remarkably each cell carries quantum
angular momentum numbers. Unfortunately some of the generators are
not hermitian, so one does not end up with a finite number of
states to satisfy the entropy bound.

The main difficulty to overcome is thus how to include in a
consistent way in the S-matrix Ansatz proposal the transverse
components of the gravitational interactions to get covariance and
a finite number of states at the same time.

What is not trivial in particular is the fact that that the
transverse momentum is not an independent variable. Indeed, if
$Q_{in}(\tilde{x})$ is any operator valued function of the
transverse coordinates $\tilde{x}$, one has for the incoming
transverse momentum $\Pi_{in}^a ( \tilde{x})$(here $ \partial_{tr}
= \tilde{\partial}$) \be \label{transvaction} [ \Pi_{in}^a (
\tilde{x}), Q_{in} (\tilde{y} ) ] = - i \delta^{(2)} ( \tilde{x} -
\tilde{y})
\partial_{tr}^a Q_{in} (\tilde{x})
\ee One can show that an operator satisfying such properties is
given by composing ingoing longitudinal position and momentum
operators as follows
 \be \label{correttotransverso} \Pi_{in}^a (\tilde{x}) =
p_{in} (\tilde{x})
\partial_{tr}^a x_{in} (\tilde{x})
\ee The same holds of course for the outcoming transverse momenta.

In analogy with the longitudinal components of the momentum one
would  expect
\be [ \Pi_{in}^a (\tilde{x}) ,
\Pi_{out}^b(\tilde{y})  ] = - i \delta^{ab} \partial_{tr}^2
\delta^{(2)} (\tilde{x} - \tilde{y}) \ee
 On the other hand one
gets a much more complicated expression.  We have found explicitly
\be
\label{complicata} [ \Pi_{in}^a (\tilde{x}) ,
\Pi_{out}^b(\tilde{y})  ] =
 i p_{in} (\tilde{x}) p_{out}
(\tilde{y})
\partial_{tr}^a \partial_{tr}^b f (\tilde{x} - \tilde{y}) - i
\partial_{tr}^2 \delta (\tilde{x} - \tilde{y}) \partial_{tr}^b
x_{out} (\tilde{y}) \partial_{tr}^a x_{in} (\tilde{x})
\ee
 This
algebra is again non local and in addition it  contains higher
order derivatives. In particular, transverse derivatives of the
Green function $f(\tilde{x} - \tilde{x}')$ suggest the presence of
leading order correction terms to the eikonal approximation.

These corrections have been analyzed in detail \cite{acv} and it
has been shown that they are related to gravitational emission
(one loop quantum corrections and two loops classical
corrections), not really to an external metric like the leading
order shift $f( \tilde{x}, \tilde{x }')$. They represent a sort of
``refraction" effect in the transverse plane and measure how much
the angle that a geodesic forms with $x^+ =0$ for instance changes
when the hypersurface $x^+ = 0$ is crossed. What is not trivial is
the fact that there might be points in the transverse place with
suffer a discontinuity produced by the shock {\it but} no
refraction and/or viceversa.

Therefore, when including second order derivatives of the $X^\mu$
embedding functions of our horizon-membrane (which have been
always neglected up to now), it is reasonable to expect then that
the relations (\ref{xinxout}) do not hold any more even before
including the transverse momenta. Unfortunately it appears not
obvious at all how to modify them.

It is interesting however to figure out, at least from a
qualitative point of view, what happens when one considers a more
general situation. Up to this moment, as said, second order
derivatives of  the embedding functions $X^{\mu}$ have been
neglected. There are however important differences which could be
relevant for the microstate counting once included.

First we want however to point out some differences between 't
Hooft and standard ``membrane paradigm" approaches.

To start 't Hooft is considering the embedding of the two
dimensional surface, the spacelike horizon, into four dimensional
target spacetime. The spacelike horizon is the intersection point
of past and future horizons and it is expected to be a very
complicated curved sub-manifold because of back-reaction effects.
As we have seen one chooses a static gauge for the transverse
coordinates $x^1 , x^2$. Normally, however, the static gauge is
imposed on $x^0, x^3$. In principle a double Wick rotation should
bring us to the standard situation, as suggested by 't Hooft. In
the latter case one imposes the static gauge \footnote{I thank
Shmuel Elitzur for discussions on these topics.} on the
world-volume coordinates and the transverse ones are expected to
be the analogue of the $x^{\pm} $ we have been discussing once the
rotations have been performed. This is similar to what happens in
recent D-brane models, where the non commutativity is in the
transverse directions. We imagine therefore to have performed such
Wick rotations and be in the standard case.

Secondly, in the usual membrane paradigm and its ``quantum
evolutum" stretched horizon one has in mind a $2 + 1$ membrane
embedded in $3+1$ spacetime, while here the spacelike horizon is
{\it two} dimensional. This is due to the definition
(\ref{momenti}) of incoming and outgoing momenta where one {\it
restricts} to $x^{\pm} = 0$ respectively. Instead of codimension
one we have codimension two and this will be relevant when
considering actions with second order derivatives of the embedding
functions $X^\mu$.

In general, if one requires reparametrization invariance such an
action \footnote{We adopt the following conventions and notations:
the target spacetime is d+2 dimensional with metric $g_{\mu \nu} =
\eta_{\mu \nu}$ with $\mu, \nu=1,...d+2$. $h_{ij}$ is the metric
of the $d+1$ dimensional worldvolume $W$, with $i,j=1,...,d+1$. We
split then the worldvolume coordinates $\xi^i$ as follows $\xi^i =
(\tau, \sigma^a)$, i.e. time-spatial coordinates, with
$a,b=1,...d.$ We denote with $\Sigma$ the spacelike surface
spanned by the $\sigma^a$ coordinates and with $q_{ab}$ the
induced metric on $\Sigma$. } can only depend of the induced
metric on the worldvolume
 $h_{ij} = g_{\mu \nu} \partial_i
X^\mu
\partial_j X^\nu $ and the extrinsic curvature $K_{ij} = n_{\mu ;
\nu}
\partial_i X^\mu
\partial_j X^\nu$. Therefore one has an expansion of operators
 constructed out of
these two quantities and their derivatives. Similar models have
indeed already been considered also in the case of black holes
physics \cite{michele}.(For a recent though slightly different
proposal \cite{parameter}).

Having in mind the S-matrix Ansatz, we now simply want to analyze
the momenta conijugate to the embedding functions $X^\mu (\xi)$.

We start with the Nambu Goto action, which by dimensional analysis
turns out to be the leading order term in a candidate effective
action and is supposed to emerge from the S-matrix Ansatz when
neglecting indeed second order derivatives.

Before doing this notice that if the codimension $\tilde{d}$ of
the worldvolume with respect to the bulk (i.e. the target space in
which the worldvolume is embedded) is greater than one, there will
be $\tilde{d}$ spacelike vector field $n^r$ ,
($r=1,...,\tilde{d}$), normal to the worldvolume. As a consequence
there will be $\tilde{d}$ extrinsic curvature tensors $K_{ij}^r$,
one along each normal. This will clearly affect the counting of
the horizon degrees of freedom, since more orientations are now
possible. It also in general fits with the holographic principle,
where the bulk-boundary codimension can change the holographic map
rule.

As an example consider the model quadratic in the mean extrinsic
curvature \cite{rigid1}, \cite{rigid2} with action \be S= \int_W
d^{d+1} \xi \sqrt{-h} K^2 \ee It has been used in the context of
rigid strings and its action can also be rewritten as
 \be
 \int \sqrt{-h} K_j^{i,r} K_i^{j,r} = \int \sqrt{-h} h^{ij} \partial_i
 W_{\mu \nu} \partial_j W_{\mu \nu} = \int \sqrt{-h} h^{ij}
 \nabla_i n_r \nabla_j n_r
 \ee
with $\partial_i \partial_j X^\mu = \Gamma_{ij}^k \partial_k X^\mu
+ K_{ij}^r n_r^\mu $.
  Notice the appearance of the orientation tensor $W^{\mu \nu}$
   which has been indeed proposed by 't Hooft as a
candidate to describe the horizon algebra. But in 't Hooft's
picture the codimension is two, so these objects should carry an
additional index corresponding to two different normals. This
might in principle give a more refined microstate counting. It
appears however including second order derivatives of the
embedding functions as showed. These contributions are indeed
neglected in 't Hooft's derivation and we suspect that this is one
of the reasons of the unbounded number of states given in the
resulting algebra.

We {\it insist} on this point since the horizon orientation is
important with respect to the {\it observer} we are considering. A
concrete example is given in the $2+1$ dimensional case, where to
make the algebra covariant one performs at some point a coordinate
transformation with a jacobian whose signs tells us if one is
inside or outside the horizon \cite{tre}.(See also \cite{sebas})

Let us finally include second order derivatives of the embedding
functions. For simplicity we choose target flat spacetime which
fits nicely also with the splitting (\ref{splitting}) of the
S-matrix for the near horizon region we are interested.

We consider first the Nambu-Goto action. One has
 \be
\label{nambugoto} S_{NG}= T \int_{W} d^{d+1} \xi \sqrt{-h} \ee
where $h_{ij} = \eta_{\mu \nu}
\partial_i X^\mu
\partial_j X^\nu $ is the induced metric on the worldvolume and
$ X^\mu ( \xi^i) = X^\mu ( \tau , \sigma^a )$ are the embedding
functions with the worldvolume splitting of coordinates $(\tau,
\sigma^a)$.

Varying the action w.r.t. the embedding functions $X^\mu$ one gets
\be \label{ngeom} \delta S_{NG} = T \int_W d^{d+1} \xi \nabla_j
\left(  \sqrt{ -h} h^{ij} \partial_i X^\mu \delta X_\mu   \right)
-T \int_W d^{d+1} \xi \nabla_j \left( \sqrt{-h} h^{ij}
\partial_i X^\mu \right) \delta X_\mu
 \ee
where the first term on the r.h.s is a total derivative. One is
then invited to apply Stokes theorem \footnote{Recall that for a
vector $B^i$ Stokes theorems goes as follows
\be \label{stokes}
\int_W d^{d+1} \xi \nabla_i (\sqrt{-h} B^i) = - \int_\Sigma d^d
\sigma \sqrt{-q} \lambda_i B^i
 \ee
where $\lambda_i$ is a timelike vector normal to $\Sigma$ and
$q_{ab} = h_{ij} \partial_a X^i
\partial_b X^j$ with $\partial_a =
\partial / \partial \sigma^a $
 is the induced metric on $\Sigma$}
 to get
\be \label{finalNG} \delta S_{NG} = -T \int_\Sigma d^d \sigma
\sqrt{-q} \lambda_j h^{ij} \partial_i X^\mu \delta X_\mu - T
\int_W d^{d+1} \xi \nabla_j \left( \sqrt{-h} h^{ij} \partial_i
X^\mu \right) \delta X_\mu
 \ee
On the other hand one has for a generic action depending only on
first order derivatives of $X^\mu$
 \be \label{findp} \delta S = \int_W d^{d+1}
\xi \sqrt{-h} \frac{\delta S}{\delta X_\mu } \delta X^\mu +
\int_{\Sigma} d^d \sigma p_\mu \delta X^\mu
 \ee
  where $p^\mu$ is
the conjugated momentum to $X^\mu$.

Comparing (\ref{finalNG}) with (\ref{findp}), it follows
 \be p^\mu
= -T \sqrt{ -q} \lambda^i
\partial_i X^\mu \ee
One sees that in this case the momentum has {\it only} tangential
components to the worldvolume.

We now consider the case in which second order derivatives of the
embedding functions $X^\mu$ are present. Recall from standard
classical mechanics that if we have an action like
\cite{Whittaker}
\be S = \int_M dt L(q, \dot{q} , \ddot{q})
\ee
one will have not only a momentum $p$ conjugated to $q$ but also a
momentum $\Pi$ conjugated to $\dot{q}$. They are easily found to
be \be \Pi = \frac{\delta L}{\delta \ddot{q}},
 p= \frac{\delta L}{ \delta \dot{q}} -
\partial_t (\frac{\delta L }{ \delta \ddot{q}})
\ee

Consider therefore the most generic worldvolume action containing
second order derivatives of the embedding functions. As remarked
before to assure reparametrization invariance it has to be
 of the
form
 \be \label{genK} S = \int_W d^{d+1} \xi \sqrt{-h} L (h_{ij} , K_{ij})
 \ee
We have assumed for simplicity codimension one, otherwise, as
pointed out before, the extrinsic curvature will carry additional
indices. The general variation is given by \be \delta S = \int_W
d^{d+1} \xi\ \sqrt{-h} \frac{\delta L}{\delta X^\mu} \delta X_\mu
+ \int_\Sigma \Pi^\mu \delta \dot{X}_\mu + p^\mu \delta X_\mu \ee
where one reads from the boundary term the momenta $p^\mu$ and
$\Pi^\mu$.  On the other hand, from the action (\ref{genK}) one
gets (integrating by parts and using $\delta K_{ij}= - n^\mu
\nabla_i \nabla_j \delta X_\mu$)
 \be
\label{variazioneFINALE} \delta S = \int_W d^{d+1} \xi \sqrt{-h}
\nabla_j S^j + bulk-e.o.m \ee with \be S^j = \left( L h^{ij}
\partial_i X^\mu -2 \frac{\delta L}{ \delta h_{ij}} \partial_i
X^\mu + \nabla_i (\frac{\delta L}{\delta K_{ij}} n^\mu )  \right)
\delta X_\mu - \frac{\delta L}{ \delta K_{ji} } n^\mu \nabla_i
\delta X_\mu
 \ee
where $n^\mu$ is a spacelike unit normal to the worldvolume. Once
again we can then apply Stokes theorem and find explicit
expressions for $\pi^\mu$ and $\Pi^\mu$ (making manifest the
dependence on $\dot{X^\mu}$). We stop here however to observe that
that the momentum $p^\mu$, contrary to the Nambu-Goto situation,
has now {\it also} components which are ${\it not}$ tangential to
the worldvolume.

When deriving the algebra (\ref{orienalgebra}) these components,
which as shown come about when including second order derivatives,
have not been considered and this might be one of the reasons for
which one does not get a finite number of states \cite{mio}. There
are also the $\Pi^\mu$ components to be taken into account and we
do not have a clear interpretation of their role at the moment.

\section{Concluding remarks}

In this paper we have revisited 't Hooft S-matrix Ansatz for
quantum black holes. We have analyzed the whole proposal from
different perspectives which we now summarize.

We started in Section $3$ by considering the effects of the
gravitational backreaction on the Hawking radiation. We have seen
that that there is shift of the phase of the outgoing wavepackets
which has physical effects. Of course if outgoing particles were
eigenstates with respect to Kruskal momentum, this effect would be
irrelevant. However this is {\it not} the case in general and
therefore the quantum state {\it changes}.

If one then asks questions concerning the nature of the Hawking
radiation, these effects do not change anything at long times
(except very short and small fluctuations during the evaporation
of the black hole which we have computed for a spherically
symmetric shell) : therefore one still has thermal spectrum and
gaussian correlators, i.e. total absence of correlations among
wave-packets. Indeed Hawking radiation simply
 originates now in
a different region that it would have if particles had not been
send into the black hole.

The S-matrix Ansatz suggests that instead of a density matrix,
which always gives thermal spectrum when tracing over an inner
region of space-time, one should try on the contrary to construct
such a detailed and precise theory able to {\it produce} this
S-matrix itself. Remember that the S-Matrix is {\it assumed} and
one looks at the back-reaction effects on it. This is of course
not an easy task; quoting \cite{uglum} ''...if the microscopic
laws were known, computing an S matrix would , according to this
view, be as daunting as computing the scattering of laser light
from a chunk of black coal".

In Section $4$ we have considered the map bulk boundary fields and
applied it to have a clear derivation non commutative algebras
between ingoing and outgoing bulk fields just using 't Hooft's
results. These non-commutative structures seem to be particularly
relevant in the context of black hole physics since they represent
first of all a natural manifestation of the complementarity
principle \cite{gerard}, \cite{uglum} and secondly should help in
reducing the degeneracy of states of the near horizon region.

In Section $5$ we have made some analogies with liquid droplets
and two dimensional electrostatic. We have seen that despite some
analogies with droplet physics and the intriguing fact that for
high temperature non compact clusters have to be included
(implying more coarse graining and therefore more information
storing) the horizon fluctuations according to the S-matrix
necessitate a membrane-like picture. We are still exploring the
analogy with two dimensional electromagnetism to see if one can
make more contact with stringy aspects. In particular we are
trying to understand what kind of ``electric circuit" should be
used for contributions beyond the eikonal.

In Section $6$ we have pointed out some of the difficulties
concerning the inclusion of the transverse momenta. It is at this
level that all the non-linearities of the gravity are expected to
play a role. We have seen that the inclusion of higher derivatives
could be in principle a good tool to guess the right horizon
algebra. The inclusion of transverse momenta is supposed to
produce then a granular structure on the horizon. This should come
out from the horizon algebra itself even if not in an easy way of
course.

We are therefore first trying to understand the connection between
the role of the transverse degrees of freedom and the interesting
result that the entropy rate production of a black hole behaves
like the one of a {\it one} dimensional system
\cite{onedimensional}, displaying dimensional reduction in analogy
with the entropy area law for black holes. \footnote{We thank
Jacob Bekenstein for introducing us to the ideas expressed in
\cite{onedimensional} and Gerard 't Hooft for the suggestion just
mentioned. }

\begin{center} {\bf Acknowledgments}
\end{center}

I would like to sincerely and strongly thank Gerard 't Hooft for
many detailed explanations, interesting suggestions and above all
for his unique style. This work is supported by the European Marie
Curie Intra-European Fellowship (contract MEIF-CT-2003-502412).

\end{document}